\documentclass[superscriptaddress,preprintnumbers,nofootinbib,showpacs]{revtex4}
\usepackage{amsmath}
\usepackage{bm}
\usepackage{latexsym}

\newcounter{fig}

\begin{document}


\rightline{ITP-UU-14/02, SPIN-14/02}

\vskip .5cm

\title{\Large\bf A photon propagator on de Sitter in covariant gauges}
\bigskip

\author{\large Silvije Domazet$^{{\rm a}*}$ and Tomislav Prokopec$^{{\rm b}}$}
\email[]{sdomazet@irb.hr;  T.Prokopec@uu.nl}


\affiliation{$^{\rm a}$ Theoretical Physics Division, 
  Rudjer Bo\v{s}kovi\'{c} Institute, \\
   P.O.Box 180, HR-10002 Zagreb, Croatia}
\affiliation{$^{\rm b}$ Institute for Theoretical
Physics (ITP) \& Spinoza Institute, Utrecht University, Postbus
80195, 3508 TD Utrecht, The Netherlands}

\begin{abstract}

\begin{center}{\bf Abstract}\end{center}

 We construct a de Sitter invariant photon propagator in general 
covariant gauges. Our result is a natural generalization of the 
Allen-Jacobson photon propagator in Feynman gauge.
Our propagator reproduces the correct response to a point static charge
and the one-loop electromagnetic stress-energy tensor, strongly suggesting 
that it is suitable for perturbative calculations on de Sitter. 

\end{abstract}


\maketitle

\section{Introduction}
\label{Introduction}

 In their classic paper Allen and Jacobson~\cite{Allen:1986}
have obtained the photon 
propagator in de Sitter invariant gauges. They have considered 
both massless and massive photons, in which case the propagator must be transverse
on both legs (a minor error was subsequently corrected 
by Tsamis and Woodard in~\cite{Tsamis:2006gj}). 
In this paper we generalize the result of Allen and Jacobson and derive
a photon propagator in general covariant gauges, in which 
the photon propagator shows explicit dependence on the gauge parameter 
$\xi\; (-\infty<\xi<\infty)$.
One can use this propagator to perform loop calculations of various 
quantities and investigate gauge independence of physical observables 
by studying whether they depend on $\xi$.

 While the transverse Allen-Jacobson propagator was shown to give 
physically reasonable answers (see {\it e.g.} the two-loop stess-energy 
calculations~\cite{Prokopec:2008gw,Prokopec:2007ak,Prokopec:2006ue}
performed with the (corrected) transverse propagator from Ref.~\cite{Tsamis:2006gj}),
it was argued in~\cite{Kahya:2005kj} that the 
Allen-Jacobson massless photon propagator in Feynman gauge does not 
give an acceptable one-loop self-mass for a charged scalar field
on de Sitter. The authors Kahya and Woodard attribute this 
unphysical behavior to the insistence of Allen and Jacobson to respect 
de Sitter symmetry. When a simple non-invariant propagator from 
Ref.~\cite{Kahya:2005kj} was used, 
one gets physically acceptable answers for the one-loop scalar self-mass.
Yet, the two propagators were calculated in different gauges, and 
it remained unclear how the choice of gauge affects the self-mass, and 
the resulting (in principle measurable) change in the field amplitude. 
Of course, a photon propagator in general covariant gauges presented here
allows to explicitly investigate  how gauge dependence enters into quantities
such as self-energies.

Vector propagators on de Sitter are useful for studying various (perturbative) 
quantum properties of theories on de Sitter space. Up to now, quantum loop effects
on de Sitter have been investigated in scalar electrodynamics in
Refs.~\cite{Prokopec:2002uw,Prokopec:2002jn,Prokopec:2003bx,Prokopec:2003iu,Prokopec:2003tm,Prokopec:2004au,Akhmedov:2010ah,Leonard:2012si,Leonard:2012ex,Glavan:2013jca}
and in~\cite{Prokopec:2008gw,Prokopec:2007ak,Prokopec:2006ue,Kahya:2005kj},
where photon and scalar field mass generation has been studied,
as well as quantum backreaction from created inflationary photons
and charged scalars~\cite{Prokopec:2008gw,Prokopec:2007ak,Prokopec:2006ue}.
Recently, a study the one-loop quantum gravitational effects on 
the photon vacuum polarization on de Sitter has appeared
in~\cite{Leonard:2013xsa,Leonard:2012si,Leonard:2012ex,Leonard:2012fs}, 
albeit gauge dependence of the results has not yet been properly addressed. 
Since de Sitter is the model space for inflation, understanding the physics of
de Sitter is of crucial importance for understanding inflationary models
(in which the Hubble parameter is an adiabatic function of time).
 An important problem of de Sitter space is known as linearization
instability~\cite{Deser:1973zza,Abbott:1981ff,Higuchi:1991tk,Tsamis:1992xa}.
Even though de Sitter space it non-compact, spatial sections of de Sitter in 
global coordinates (with positively curved spatial sections) 
are compact, and hence for these sections the usual considerations
apply, according to which no net charge can be placed on a compact space.
This follows immediately from the Gauss's law
$\nabla_\mu F^{\mu\nu}=(-g)^{-1/2}\partial_\mu [(-g)^{1/2}F^{\mu\nu}]=J^\nu$
which, when written in the integral form, implies
\begin{equation}
  \int_{\Sigma_t} d^3x \sqrt{\gamma}\nabla_i F^{i 0} 
     =  \int_{\Sigma_t} d^3x \partial_i [\sqrt{\gamma}F^{i 0}] 
     = \int_{\Sigma_t} d^3x \sqrt{\gamma} J^0 \equiv Q(t)
\,,
\label{Gauss's law}
\end{equation}
where $Q(t)$ denotes the total charge on the spatial section $\Sigma_t$ 
at time $t$ and $\gamma$ is the determinant of the spatial part of the metric
(here it is assumed that one works in coordinates in which
$g_{0i}=0$ and $g_{00}$ is independent on spatial coordinates).
Eq.~(\ref{Gauss's law}) can be also written as, 
\begin{equation}
Q(t) =  \int_{\partial\Sigma_t} dS \sqrt{\gamma_S}\hat n_iF^{i 0} 
\,,
\label{Gauss's law:2}
\end{equation}
where $n_i$ is the unit vector orthogonal to the boundary surface
$S=\partial\Sigma_t$ and $\gamma_S$ is the determinant of the induced two dimensional
metric on $\partial\Sigma_t$.
Since spatial sections of a compact space have no boundary, the integral 
in~(\ref{Gauss's law:2}) must vanish, and therefore no net charge can be placed
on a compact surface. De Sitter space is non-compact however, and hence 
strictly speaking this consideration applies only to global sections 
of de Sitter space, which are compact. In this paper we consider 
the response to a point charge on flat sections of de Sitter
space, which are non-compact, and hence the above restriction does not apply. 

Before embarking on calculations in de Sitter space-time we shall first lay out
a calculation of the response to a static point charge on
Minkowski space. The (Keldysh) photon propagator on Minkowski background
in covariant gauges and in $D=4$ space-time dimensions is given by,
\begin{equation}
 \imath [_\mu\Delta^{ab}_\nu](x;x^\prime)
  =\eta_{\mu\nu}\imath\Delta^{ab}_0(x;x^\prime)
     - \imath(1-\xi)\partial_\mu\partial_\nu\int d^4x^{\prime\prime}
             \Delta^{ac}_0(x;x^{\prime\prime})(\sigma^3)^{cd}
                  \Delta^{db}_0(x^{\prime\prime};x^\prime)
\,,
\label{photon propagator:Minkowski}
\end{equation}
where $a,b,c,d$ (a summation over the repeated indices is assumed),
$\sigma^3={\rm diag}(1,-1)$ and 
\begin{equation}
 \imath \Delta_0^{ab}(x;x^\prime)
  =\frac{1}{4\pi^2}\frac{1}{[\Delta x^{ab}(x;x^\prime)]^2}
\,,
\label{photon propagator:Minkowski2}
\end{equation}
is the massless scalar propagator on Minkowski in four space-time dimensions and
\begin{eqnarray}
 [\Delta x^{++}(x;x^\prime)]^2 
     &=& -(|t\!-\!t'|\!-\imath \epsilon)^2+\|\vec x\!-\!\vec x^{\,\prime}\|^2
 \,,\qquad
 [\Delta x^{+-}(x;x^\prime)]^2 
        = -(t\!-\!t'+\imath \epsilon)^2+\|\vec x\!-\!\vec x^{\,\prime}\|^2
\label{invariant distance: Mink:1}
\end{eqnarray}
\begin{eqnarray}
 [\Delta x^{--}(x;x^\prime)]^2 
     &=& -(|t\!-\!t'|\!+\imath \epsilon)^2+\|\vec x\!-\!\vec x^{\,\prime}\|^2
 \,,\qquad
 [\Delta x^{-+}(x;x^\prime)]^2 
         = -(t\!-\!t'-\imath \epsilon)^2+\|\vec x\!-\!\vec x^{\,\prime}\|^2
 \,.
\label{invariant distance: Mink:2}
\end{eqnarray}
The $(++)$ and $(--)$ components of the Keldysh propagator correspond to
the Feynman and anti-Feynman propagator, respectively,
and the  $(+-)$ and $(-+)$ components
are the negative and positive frequency 
Wightman functions, respectively, which are 
useful {\it e.g.} for non-equilibrium problems of statistical physics. 

The electromagnetic potential for a static point charge can be obtained
from the formula,
\begin{equation}
 A_\mu(x)=\int d^4x^\prime [_{\mu}\Delta^{\rm ret}_\nu](x;x^\prime)j^\nu(x^\prime)
\,,
\label{response on point chargea}
\end{equation}
where the current density is given by
\begin{equation}
 j^\nu = \delta^\nu_{\;0}e\delta^3(\vec x\,)
\,,
\label{current point chargea}
\end{equation}
and $[_{\mu}\Delta^{\rm ret}_\nu]$ is the retarded photon propagator given by
\begin{eqnarray}
  [_{\mu}\Delta^{\rm ret}_\nu](x;x^\prime)
    &=& [_{\mu}\Delta^{++}_\nu](x;x^\prime)
    - [_{\mu}\Delta^{+-}_\nu](x;x^\prime)
\nonumber\\
    &=&\eta_{\mu\nu}\left(\Delta_0^{++}(x;x^\prime)
                  - \Delta_0^{+-}(x;x^\prime)\right)
\nonumber\\
 &&-(1\!-\!\xi)\partial_\mu\partial_\nu \int d^4x^{\prime\prime}
        \left(\Delta_0^{++}(x;x^{\prime\prime})
                  - \Delta_0^{+-}(x;x^{\prime\prime})\right)
     \left(\Delta_0^{++}(x^{\prime\prime};x^\prime)
                  - \Delta_0^{+-}(x^{\prime\prime};x^\prime)\right)
\,,
\label{retarder propagator:Minkowskia}
\end{eqnarray}
where
\begin{eqnarray}
\Delta_0^{++}(x;x^\prime) - \Delta_0^{+-}(x;x^\prime)
  &=& -\frac{1}{2\pi}\theta(\eta\!-\!\eta^\prime)
   \delta\left((\eta\!-\!\eta^\prime)^2-\|\vec x\!-\!\vec x^{\,\prime}\|^2\right)
\nonumber\\
  &=& -\frac{1}{4\pi}\theta(\eta\!-\!\eta^\prime)
   \frac{\delta\left(\eta\!-\!\eta^\prime-\|\vec x\!-\!\vec x^{\,\prime}\|\right)}
           {\|\vec x\!-\!\vec x^{\,\prime}\|}
\,.
\label{retarder propagator:Minkowski:2}
\end{eqnarray}
Inserting this into~(\ref{retarder propagator:Minkowskia}) yields,
\begin{eqnarray}
  [_{\mu}\Delta^{\rm ret}_\nu](x;x^\prime)
    &=&-\frac{\eta_{\mu\nu}}{4\pi}
 \frac{\delta\left(\eta\!-\!\eta^\prime\!-\!\|\vec x\!-\!\vec x^{\,\prime}\|\right)}
           {\|\vec x\!-\!\vec x^{\,\prime}\|}
\nonumber\\
 &&-\,\frac{1\!-\!\xi}{16\pi^2}
   \partial_\mu\partial_\nu \int d^4x^{\prime\prime}
    \frac{\delta\left(\eta\!-\!\eta^{\prime\prime}
                          \!-\!\|\vec x\!-\!\vec x^{\,\prime\prime}\|\right)}
           {\|\vec x\!-\!\vec x^{\,\prime\prime}\|}
    \frac{\delta\left(\eta^{\prime\prime}\!-\!\eta^\prime
                  \!-\!\|\vec x^{\prime\prime}\!-\!\vec x^{\,\prime}\|\right)}
           {\|\vec x^{\prime\prime}\!-\!\vec x^{\,\prime}\|}
\,.
\label{retarder propagator:Minkowski:3}
\end{eqnarray}
We can now insert this into the potential equation~(\ref{response on point chargea})
and make use of the source~(\ref{current point chargea}). This gives
\begin{eqnarray}
 A_\mu(x)=\frac{e\delta_{\mu}^0}{4\pi r}\int_{\eta_0}^\eta d\eta^\prime
            \delta\left(\eta\!-\!\eta^\prime\!-\!r\right)
-(1\!-\!\xi)\frac{e}{16\pi^2}\partial_\mu\partial_0
         \int d^3x^{\prime\prime}
 \int_{\eta_0}^\eta d\eta^{\prime\prime}
   \!\int_{\eta_0}^{\eta^{\prime\prime}}\!\! d\eta^{\prime}
    \frac{\delta\left(\eta\!-\!\eta^{\prime\prime}
                          \!-\!\|\vec x\!-\!\vec x^{\,\prime\prime}\,\|\right)}
           {\|\vec x\!-\!\vec x^{\,\prime\prime}\|}
\frac{\delta\left(\eta^{\prime\prime}\!-\!\eta^\prime\!-\!r^{\prime\prime}
           \right)}{r^{\prime\prime}}
\,,\quad
\label{response to point charge}
\end{eqnarray}
where $r=\|\vec x\|$, $r^\prime=\|\vec x^{\,\prime}\|$
and $r^{\prime\prime}=\|\vec x^{\,\prime\prime}\|$. 
Performing the necessary integrations
(see Appendix~\ref{The photon propagator on Minkowski background}) results in,
\begin{eqnarray}
 A_\mu(x)&=&\frac{e\delta_{\mu}^0}{4\pi r}\theta(\Delta\eta_0\!-\!r)
\label{response to point charge:6}
\\
 &&\hskip .cm
-\,(1\!-\!\xi)\frac{e}{4\pi}\partial_\mu\frac{1}{r}
   \int_{0}^{\Delta \eta_0} d\Delta\eta
 \Big\{\theta(\Delta\eta\!-\!r)
-\theta(r)\theta(r\!-\!2\Delta\eta\!+\!\Delta\eta_0)
   +2\theta(r\!-\!\Delta \eta)-\theta(r\!-\!\Delta \eta_0)
\Big\}
\,,
\nonumber
\end{eqnarray}
where $\Delta\eta^\prime = \eta-\eta^\prime$ and $\Delta\eta_0=\eta-\eta_0$.
This result can be written as
\begin{equation}
 A_\mu(x) = \delta_{\mu}^0 {\cal A}_0(x) + \partial_\mu \Lambda(x)
\label{response to point charge:7}
\,,
\end{equation}
where
\begin{eqnarray}
{\cal A}_0(x) = \frac{e}{4\pi r}\theta(\Delta \eta_0-r)
\,,\qquad
\Lambda(x) =-\,(1\!-\!\xi)\frac{e}{8\pi}\theta(\Delta \eta_0-r)
                \Big(1+\frac{\Delta\eta_0}{r}\Big)
\,.
\label{response to point charge:7bb}
\end{eqnarray}
From Eq.~(\ref{response to point charge:7}) it is clear that $\Lambda$ in~(\ref{response to point charge:7bb})
is a pure gauge contribution, and does not affect
physics. On the other hand, ${\cal A}_0$ contributes to the physical electric field
as
\begin{equation}
\vec  E= -\nabla A_0 + \partial_0 \vec A = -\nabla {\cal A}_0
= \frac{e}{4\pi}\frac{\vec r}{r^3}\left\{\theta(\Delta\eta_0-r)
                     +r\delta(\Delta\eta_0-r)
                \right\}
\,,\qquad \vec B = \nabla\times \vec A = 0
\,.
\end{equation}
 The time dependence
(on $\Delta \eta_0$) in the above expressions arises from unphysical
initial conditions. Namely, the state at $\eta=\eta_0$ is chosen such as if there was no charge
at $\eta<\eta_0$ which generates a light-cone starting at $\eta_0$ and propagating along
$\Delta \eta_0=r$. Of course, one cannot create or destroy a charge, hence these initial
conditions are unphysical, and one gets the physical answer by
sending $\eta_0\rightarrow -\infty$, in which case all time dependence
(in the physical part of the electric field) disappears and we get
\begin{equation}
 {\cal A}_0 = \frac{e}{4\pi r}
\,,\qquad \vec E=\frac{e}{4\pi}\frac{\vec r}{r^3}
\,,\qquad \vec B= 0
\,,
\label{physical fields}
\end{equation}
which is (obviously) the correct answer.


\section{Calculating the propagator}
\label{Propagator}

After giving a short account of the calculation in Minkowski space
we now turn our attention to
solving the equivalent problem in de Sitter space. The relevant curved space
action is given by
\begin{eqnarray}
S_{\rm EM}=\int d^4x\sqrt{-g}
 \bigg[-\frac14F_{\mu\nu}F^{\mu\nu}-\frac{1}{2\xi}(\nabla_{\mu}A^{\mu})^2\bigg]
\,,
\label{photon action}
\end{eqnarray}
where $F_{\mu\nu}=\nabla_\mu A_\nu -\nabla_\nu A_\mu
             = \partial_\mu A_\nu -\partial_\nu A_\mu$ is the field strength
and the last term is added as a covariant gauge fixing term ($\xi\in(-\infty,+\infty)$).
By varying the action~(\ref{photon action})
 one gets that the photon field satisfies the equation,
\begin{equation}
 L^{\mu\nu}A_{\nu}=0
\,.
\label{eom:photon field}
\end{equation}
where 
\begin{eqnarray}
\label{photoper}
   L^{\mu\nu}=g^{\mu\nu}\Box-\nabla^\nu \nabla^\mu
    +\frac{1}{\xi}\nabla^{\mu}\nabla^\nu
  =g^{\mu\nu}\Box
    -\Big(1-\frac{1}{\xi}\Big)\nabla^{\mu}\nabla^\nu -R^{\mu\nu}
\,.
\label{photon operator Lmn}
\end{eqnarray}
Here $R^{\mu\nu}$ denotes the Ricci tensor which in de Sitter space equals to 
$H^2(D-1)g^{\mu\nu}$. The equation satisfied by the photon propagator is then
\begin{eqnarray}
L^{\mu\nu}(x)\imath [_{\nu}\Delta_{\alpha}](x;x^\prime)
 = \bigg[g^{\mu\nu}\Box-\nabla^\nu\nabla^\mu
          +\frac{1}{\xi}\nabla^{\mu}\nabla^\nu
          \bigg]\imath [_{\nu}\Delta_{\alpha}](x;x^\prime)
= (\partial^\mu\partial^{\prime}_{\alpha}y)
         \frac{\imath\delta^D(x\!-\!x')}{\sqrt{-g}(-2H^2)}
\,,
\label{photon propagator:eom}
\end{eqnarray}
plus the equation 
$L^{\beta\alpha}(x^\prime)\imath [_{\nu}\Delta_{\alpha}](x;x^\prime)
=(\partial^{\prime\beta}\partial_\nu y)[\imath\delta^D(x\!-\!x')/(-2H^2\sqrt{-g})]$,
which is automatically satisfied when the following exchange symmetry is imposed,
\begin{equation}
\imath[_{\nu}\Delta_{\alpha}](x;x^\prime) = \imath[_{\alpha}\Delta_{\nu}](x^\prime;x)
\,.
\label{exchange symmetry}
\end{equation}
The factor $-2H^2$ in the denominator on the right hand side 
of~(\ref{photon propagator:eom}) is chosen to ensure the correct normalization of
the propagator. Indeed, because of the delta function, 
$(\partial^\mu\partial^{\prime}_{\alpha}y)$
on the right hand side of Eq.~(\ref{photon propagator:eom}) 
can be replaced by $-2H^2 \delta^\mu_{\;\alpha}$,
such that in the limit when $H\rightarrow 0$  
the propagator~(\ref{photon propagator:eom}) reduces to
its Minkowski counterpart~(\ref{photon propagator:Minkowski}), as it should.
Our {\it Ansatz} for the propagator consists of two de Sitter invariant
tensor structures, each multiplying a de Sitter invariant scalar structure function,
so that
\begin{eqnarray}
\label{photprop}
  \imath [_{\nu}\Delta_{\alpha}](x;x^\prime)
    &=& (\partial_{\nu}\partial^\prime_{\alpha} y)\times f_1(y)
    + (\partial_{\nu}y)\times(\partial^\prime_{\alpha} y)\times f_2(y)
\nonumber\\
    &=& (\partial_{\nu}\partial^\prime_{\alpha} y)\times A_1(y)
    + \partial_{\nu}\partial^\prime_{\alpha} A_2(y)
\,,
\label{propagator: tensor structures a}
\end{eqnarray}
where
\begin{equation}
y(x;x^\prime) \equiv y^{++}(x;x^\prime)
\,,\qquad
y^{ab}(x;x^\prime) = aa^\prime H^2[\Delta x^{ab}(x;x^\prime)]^2
\qquad (a,b=+,-)
\,,
\label{y:definition}
\end{equation}
where $[\Delta x^{ab}(x;x^\prime)]^2$ are given in 
Eqs.~(\ref{invariant distance: Mink:1}--\ref{invariant distance: Mink:2}).
The latter form in~(\ref{propagator: tensor structures a}) is
motivated by the tensor structure of the photon propagator on Minkowski space,
and as we will see below it can be used to significantly simplify our equations for
the scalar structure functions.

\subsection{Solving the photon equation on de Sitter}
\label{Solving the photon equation on de Sitter}

 In Appendix~\ref{The photon tensor structures} we have shown how the
photon operator~(\ref{photon operator Lmn}) acts on
the propagator in the equation of motion~(\ref{photon propagator:eom}),
when the propagator is represented in terms of de Sitter invariant
tensor structures (more precisely bi-vectors)
and scalar structure functions $f_1(y)$ and $f_2(y)$
as in Eq.~(\ref{propagator: tensor structures a}).
Since the two tensor structures
in appendix~(\ref{L on propagator:tensor structure}) are mutually independent,
Eq.~(\ref{L on propagator:tensor structure}) implies the
following two scalar equations,
\begin{eqnarray}
  && \hskip -0.5cm
  (4y\!-\!y^2)f_1^{\prime\prime}
       +\Big(D\!-\!1\!+\!\frac{1}{\xi}\Big)(2\!-\!y)f_1^{\prime}-\frac{D}{\xi}f_1
 -\Big(1\!-\!\frac{1}{\xi}\Big)(4y\!-\!y^2)f_2^{\prime}
         -\Big(D\!-\!1-\frac{D\!+\!1}{\xi}\Big)(2\!-\!y)f_2
= \frac{\imath\delta^D(x\!-\!x')}{H^2\sqrt{-g}(-2H^2)}
\qquad
\label{scalar equation:1}
\\
&&\hskip -0.5cm
-\Big(1\!-\!\frac{1}{\xi}\Big)(2\!-\!y)f_1^{\prime\prime}
+\Big(D\!-\!1\!-\!\frac{D\!+\!1}{\xi}\Big)f_1^{\prime}
  +\frac{4y\!-\!y^2}{\xi}f_2^{\prime\prime}
  +\Big(1\!+\!\frac{D\!+\!3}{\xi}\Big)(2\!-\!y)f_2^{\prime}
   -\Big((D\!-\!1)\!+\!\frac{D\!+\!1}{\xi}\Big)f_2 = 0
\,.
\qquad
\label{scalar equation:2}
\end{eqnarray}
It turns out that it is more convenient to represent these equations
in the $A_1-A_2$ basis,
defined in the second line of Eq.~(\ref{propagator: tensor structures a}),
which implies,
\begin{equation}
 f_1 = A_1 + A_2^\prime \,;\qquad f_2 = A_2^{\prime\prime}
\,.
\label{A1-A2 basis}
\end{equation}
The rationale for this choice will soon become apparent. Namely,
in this basis Eqs.~(\ref{scalar equation:1}--\ref{scalar equation:2}) become,
\begin{eqnarray}
&& \hskip -1cm
   (4y\!-\!y^2)A_1^{\prime\prime}
       +\Big((D\!-\!1)+\frac{1}{\xi}\Big)(2\!-\!y)A_1^{\prime}-\frac{D}{\xi}A_1
 +\frac{1}{\xi}(4y\!-\!y^2)A_2^{\prime\prime\prime}
         +\frac{D\!+\!2}{\xi}\Big)(2\!-\!y)A_2^{\prime\prime}
         - \frac{D}{\xi}A_2^\prime
\label{scalar equation:A1}
  = \frac{\imath\delta^D(x\!-\!x')}{(-2H^4)\sqrt{-g}}
\quad
\\
&& \hskip -1cm
-\Big(1-\frac{1}{\xi}\Big)(2\!-\!y)A_1^{\prime\prime}
+\Big((D\!-\!1)-\frac{D\!+\!1}{\xi}\Big)A_1^{\prime}
  +\frac{1}{\xi}(4y\!-\!y^2)A_2^{\prime\prime\prime\prime}
  +\frac{D\!+\!4}{\xi}(2\!-\!y)A_2^{\prime\prime\prime}
   -\frac{2(D\!+\!1)}{\xi}A_2^{\prime\prime} = 0
\,.
\qquad
\label{scalar equation:A2}
\end{eqnarray}
The latter equation~(\ref{scalar equation:A2}) can be easily integrated once
to give,
\begin{eqnarray}
-\Big(1-\frac{1}{\xi}\Big)(2\!-\!y)A_1^{\prime}
+\Big(D\!-\!2-\frac{D}{\xi}\Big)A_1
  +\frac{1}{\xi}(4y\!-\!y^2)A_2^{\prime\prime\prime}
  +\frac{D\!+\!2}{\xi}(2\!-\!y)A_2^{\prime\prime}
   -\frac{D}{\xi}A_2^\prime = 0
\,,
\quad
\label{scalar equation:A2b}
\end{eqnarray}
where we set the integration ($y$-independent) constant to zero.
By inserting Eq.~(\ref{scalar equation:A2b}) into Eq.~(\ref{scalar equation:A1})
we see that the equation for $A_1$ decouples from that for $A_2$.
The result is the following inhomogeneous Gauss's hypergeometric equation for $A_1$,
with the usual delta function source on the light-cone:
\begin{equation}
 \frac{1}{H^2}\Big(\Box-(D\!-\!2)H^2\Big)A_1(y)
  \equiv (4y\!-\!y^2)A_1^{\prime\prime}+ D(2\!-\!y)A_1^{\prime}-(D\!-\!2)A_1
        = \frac{\imath\delta^D(x\!-\!x')}{(-2H^4)\sqrt{-g}}
\,,
\quad
\label{scalar equation:A1b}
\end{equation}
such that the de Sitter invariant photon equation on de Sitter space
for the gauge invariant scalar structure function
reduces to that of a scalar field with a (photon) mass term given
by 
\begin{equation}
m_A^2=(D\!-\!2)H^2
\,.
\label{photon mass term}
\end{equation}
Furthermore, Eq.~(\ref{scalar equation:A1b}) is gauge independent
(any dependence on $\xi$ has dropped out). This is a very welcome feature
since it tells us that the $A_1-A_2$ basis~(\ref{A1-A2 basis})
separates the photon propagator into a gauge independent part and
a gauge dependent part.

 Requiring that at light-cone the propagator reduces to
a Hadamard form yields a unique solution to
the Feynman (time ordered) propagator in
Eq.~(\ref{scalar equation:A1b}) ({\it cf. e.g.}
the Appendix in Ref.~\cite{Prokopec:2011ms}),
\begin{equation}
\imath\Delta(x;x^\prime) \equiv A_1(y(x;x^\prime))
= -\frac{H^{D-4}}{2(4\pi)^{D/2}}
 \frac{\Gamma(\frac{D-1}{2}\!+\!\nu_D)\Gamma(\frac{D-1}{2}\!-\!\nu_D)}{\Gamma(\frac{D}{2})}
 \times{}_2F_1\Big(\frac{D\!-\!1}{2}+\nu_D,\frac{D\!-\!1}{2}-\nu_D;
                      \frac{D}{2};1\!-\!\frac{y}{4}\Big)
,
\label{A1:general solution}
\end{equation}
where
\begin{equation}
\nu_D^2=\Big(\frac{D\!-\!1}{2}\Big)^2-\frac{m^2}{H^2}
\,;\qquad \frac{m^2}{H^2} = D\!-\!2
\;\longrightarrow \; \nu_D = \frac{D\!-\!3}{2}
\,
 \label{nuD}
\end{equation}
and
\begin{equation}
y(x;x^\prime) \equiv y^{++}(x;x^\prime)
           = a(\eta)a(\eta^\prime)H^2
  \big[\!-(|\eta\!-\!\eta^\prime|-\imath \epsilon)^2
           +\|\vec x\!-\!\vec x^{\,\prime}\|^2\big]
\,.
\nonumber
\end{equation}
To get the other three elements of the $2\times2$ Keldysh propagator,
one needs to replace $y=y^{++}$ by $y^{--}$ and $y^{\pm\mp}$
in~(\ref{A1:general solution}), respectively, 
see Eq.~(\ref{y:definition}).
Since $\nu_D=(D\!-\!3)/2$, the general solution~(\ref{A1:general solution})
can be simplified to,
\begin{equation}
 A_1(y) = -\frac{H^{D-4}}{2(4\pi)^{D/2}}
 \frac{\Gamma(D\!-\!2)}{\Gamma(\frac{D}{2})}
 \times{}_2F_1\Big(D\!-\!2,1;\frac{D}{2};1-\frac{y}{4}\Big)
\,.
\label{A1:photon solution}
\end{equation}
The gauge independent part of the Keldysh propagator is then simply,
\begin{equation}
 \imath\Delta^{ab}(x;x^\prime) = -\frac{H^{D-4}}{2(4\pi)^{D/2}}
 \frac{\Gamma(D\!-\!2)}{\Gamma(\frac{D}{2})}
 \times{}_2F_1\Big(D\!-\!2,1;\frac{D}{2};1-\frac{y^{ab}(x;x^\prime)}{4}\Big)
\,;\qquad (a,b,=\pm)
\,.
\label{photon Keldysh propagator:A1}
\end{equation}
If we are interested in the behavior of $A_1$ near the light cone ($y\sim 0$),
then Eq.~(9.131.2) of Gradshteyn and Ryzhik~\cite{GradshteynRyzhik:2007} 
can be used to transform~(\ref{A1:photon solution}) into
\begin{equation}
 A_1(y)  = -\frac{H^{D-4}}{2(4\pi)^{D/2}}
\bigg[ \Gamma\Big(\frac{D\!-\!2}{2}\Big)\Big(\frac{y}{4}\Big)^{-(D\!-\!2)/2}
      \times{}_1F_0\Big(\frac{D\!-\!2}{2};\frac{y}{4}\Big)
     -\frac{\Gamma(D\!-\!2)}{\Gamma(\frac{D}{2})}
      \times{}_2F_1\Big(D\!-\!2,1;\frac{D}{2};\frac{y}{4}\Big)
\bigg]
\,.
\label{A1:photon solution:light-cone}
\end{equation}
Obviously, it is the first part of
the propagator~(\ref{A1:photon solution:light-cone}) that yields the
Hadamard behavior near the light-cone, $ A_1(y)\propto y^{-(D\!-\!2)/2}
       \propto (\Delta x^2)^{-(D\!-\!2)/2}$, where $\Delta x^2\simeq 0$.

 What remains to be done is to solve for $A_2$, which
can be done by solving the following inhomogeneous
equation~(\ref{scalar equation:A2b}),
\begin{eqnarray}
(4y\!-\!y^2)A_2^{\prime\prime\prime}
  +(D\!+\!2)(2\!-\!y)A_2^{\prime\prime}
   -D A_2^\prime
= -\big(1\!-\!\xi\big)(2\!-\!y)A_1^{\prime}+\big(D-(D\!-\!2)\xi\big)A_1
\,,
\qquad
\label{scalar equation:A2c}
\end{eqnarray}
Since we have previously established that $A_1$ is independent of the gauge parameter $\xi$,
from Eq.~(\ref{scalar equation:A2c}) we see that $A_2$ depends on $\xi$.
Next, we can integrate Eq.~(\ref{scalar equation:A2c}) once to get
\begin{eqnarray}
(4y\!-\!y^2)A_2^{\prime\prime}+D(2\!-\!y)A_2^\prime
 &=& \frac{1}{(4y\!-\!y^2)^{(D\!-\!2)/2}}
     \frac{d}{dy}\bigg[(4y\!-\!y^2)^\frac{D}{2}A_2^\prime\bigg]
\nonumber\\
&=& -(1\!-\!\xi)(2\!-\!y)A_1+\Big((D\!-\!1)-(D\!-\!3)\xi\Big)I[A_1]
  \equiv s_\xi(y)
\,.\quad
\label{eom:A2:I}
\end{eqnarray}
What this tells us is that the Green's function for $A_2$ is that of
the scalar d'Alembertian,
$\Box G(x;x^\prime) = H^2\delta^D(x-x^\prime)/\sqrt{-g}$,
which is known to have no de Sitter invariant solution
of a Hadamard form. This can be easily seen from the expression after that
first equality in~(\ref{eom:A2:I}), which can be easily integrated, to yield
(up to a constant) for the Green's function of $A_2$,
\begin{equation}
 \imath G_{A_2}(x;x^\prime)\propto \int \frac{dy}{(4y-y^2)^{D/2}}
     \propto (4y-y^2)^{1-\frac{D}{2}}
\times{}_2F_1\Big(1,2-D;2-\frac{D}{2};\frac{y}{4}\Big)
\,.
\label{Green function: scalar dAlembertian}
\end{equation}
Acting with the scalar d'Alembertian on this solution, and
taking a careful account of the $\imath \epsilon$ prescription (for the
time ordered propagator) reveals that this solution is a response to
a source $\propto \delta^D(x-x^\prime)$ located at the light-cone where $y=0$,
as it should, but also to an additional source at the antipodal point,
where $y=4$ (because of the $\propto (4-y)^{1-D/2}$ behavior
in~(\ref{Green function: scalar dAlembertian}) close to $y=4$).
Since there is no source at the antipodal point, this behavior is unphysical.
This simply means that there exists no de Sitter invariant Green's function 
of the Hadamard form that solves Eq.~(\ref{eom:A2:I}). 
One can proceed in two ways: (a)
ignore that problem and write down a de Sitter invariant form
for the solution (even though that means that a fictitious source
at the antipodal point will also contribute), or (b)
construct a proper Green's function for the problem
that respects the Hadamard form but breaks de Sitter symmetry.
In the light of this discussion, it is unclear to us how
Allen and Jacobson in~\cite{Allen:1986} could have obtained 
a photon propagator that
is both de Sitter invariant and of the Hadamard form.

 To be more concrete, choosing the de Sitter invariant option leads to
\begin{eqnarray}
\imath G_{A_2}
=\frac{2^{D-5}H^{D-2}\Gamma\Big(\frac{D-2}{2}\Big)}{\pi^{D/2}}(4y\!-\!y^2)^{1-D/2}
\times\bigg\{{}_2F_1\Big(1,2\!-\!D;2\!-\!\frac{D}{2};\frac{y}{4}\Big)-
{}_2F_1\Big(1,2\!-\!D;2\!-\!\frac{D}{2};1\!-\!\frac{y}{4}\Big)\bigg\}
\label{total solution GIA_2}
\,,
\end{eqnarray}
where in order to cancel the annoying constant term $\propto 1/(D-4)$ 
that arises when expanding the hypergeometric function 
in~(\ref{Green function: scalar dAlembertian}) around $D=4$,
we have added a second hypergeometric function
which also solves the same equation. 
This is legitimate, since the indefinite integral
in~(\ref{Green function: scalar dAlembertian}) is defined up to a constant.
The expression~(\ref{total solution GIA_2}) leads to the following $D \to 4$ limit
\begin{equation}
\imath G_{A_2}=\frac{H^2}{4\pi^2}\bigg\{\frac{1}{y}-\frac12\ln y-\frac{1}{4-y}+\frac12\ln (4-y)\bigg\}
\label{secondDis4}
\, ,
\end{equation}
which again clearly exhibits the presence of an unphysical source 
at the antipodal point at $y=4$.
As we have already mentioned, avoiding this problem
by resorting to the second option means one could take
a propagator that respects spatial translations but breaks 
the more general de Sitter symmetry~\cite{Onemli:2004mb}
\begin{eqnarray}
\imath\Delta_{\rm new}(x;x^\prime)&=&\frac{H^{D-2}}{(4\pi)^{D/2}}
\bigg\{\!-\!\sum_{n=0}^{\infty}
  \frac{1}{n\!-\!\frac{D}{2}\!+\!1}
  \frac{\Gamma\big(n\!+\!\frac{D}{2}\big)}
       {\Gamma\big(n\!+\!1\big)}\Big(\frac{y}{4}\Big)^{n-(D/2)+1}
\!-\frac{\Gamma\big(D\!-\!1\big)}{\Gamma\big(\frac{D}{2}\big)}
        \pi\cot{\Big(\frac{\pi D}{2}\Big)}
\nonumber \\
&&\hskip 1.5cm
+\,\sum_{n=1}^{\infty}\frac{1}{n}
\frac{\Gamma\big(n\!+\!D\!-\!1\big)}
     {\Gamma\big(n\!+\!\frac{D}{2}\big)}\Big(\frac{y}{4}\Big)^{n}
  +\frac{\Gamma\big(D\!-\!1\big)}{\Gamma\big(\frac{D}{2}\big)}\ln(aa')
\bigg\}
\,,
\label{IA2:photon solution dS break}
\end{eqnarray}
for which the $D \to 4$ limit is given by
\begin{eqnarray}
\imath\Delta_{\rm new}(x;x^\prime)\;\stackrel{D\to 4}{\longrightarrow}\;
\frac{H^2}{(4\pi)^2}\bigg\{\frac{4}{aa'H^2(\Delta x)^2}-2\ln\bigg(\frac{H^2(\Delta x)^2}{4}\bigg)-1\bigg\}
\label{newDis4}
\end{eqnarray}
For both choices the Minkowski limit ($H\to 0$, $a,a'\to 1$) 
gives the massless scalar propagator on Minkowski space, as it should, namely
\begin{equation}
\imath\Delta_{\rm new}(x;x^\prime)\;\stackrel{{\rm Mink}}{\longrightarrow}\;
\frac{\Gamma\big(\frac{D\!-\!2}{2}\big)}{4\pi^{D/2}}
            \frac{1}{(\Delta x^2)^{\frac{D-2}{2}}}
\label{newmink}
\, ,
\end{equation}
which for $D \to 4$ reduces to
\begin{eqnarray}
\imath\Delta_{\rm new}(x;x^\prime)\;\stackrel{D\to 4,{\rm Mink}}{\longrightarrow}\;
\frac{1}{4\pi^{2}}\frac{1}{\big(\Delta x)^2}
\label{secondmink}
\, .
\end{eqnarray}
Once equipped with the Green's function for $A_2$ we can write
the solution for $A_2$ as
\begin{equation}
A_2^{ab}(x;x')=H^2\int d^D x^{\prime\prime}G_{A_2}^{ac}(x;x^{\prime\prime})
            (\sigma^3)^{cd}s_{\xi}^{db}(x^{\prime\prime};x') \, ,
\label{A2:final}
\end{equation}
where the source is given by, 
\begin{equation}
s_{\xi}(x^{\prime\prime};x')=\{-(1\!-\!\xi)(2\!-\!y)A_1(y)
  +\big((D\!-\!1)-(D\!-\!3)\xi\big)I[A_1](y)\}(x^{\prime\prime};x^{\prime})
\, .
\label{source sxi}
\end{equation}
Having in mind the result (\ref{A1:photon solution})
we see that finding the contribution of $I[A_1]$ to the above expression amounts to
evaluating the following integral
\begin{equation}
\int dy ~{}_2F_1\Big(D\!-\!2,1;\frac{D}{2};1-\frac{y}{4}\Big)
\, .
\end{equation}
After making the substitution, $z=1-y/4$, the relevant integral is
\begin{equation}
\label{odgovor}
\int dz ~{}_2F_1\Big(a,b;c;z\Big)=\frac{c\!-\!1}{a\!-\!1}
\times\frac{{}_2F_1\Big(a\!-\!1,b\!-\!1;c\!-\!1;z\Big)-1}{b\!-\!1}
\, ,
\end{equation}
where for convenience we have added an integration constant.
This gives
\begin{eqnarray}
I[A^{db}_1(y)]&=&\frac{H^{D-4}}{(4\pi)^{D/2}}
 \frac{\Gamma(D\!-\!1)}{\Gamma(\frac{D}{2})}
 \times\frac{1}{D\!-\!3}{\rm lim}_{b\to 1}
 \Bigg[\frac{{}_2F_1\Big(D\!-\!3,b\!-\!1;\frac{D-2}{2};1\!-\!\frac{y^{db}}{4}\Big)-1}
             {b\!-\!1}\Bigg]
\nonumber\\
&=&\frac{2H^{D-4}}{(4\pi)^{D/2}}\sum_{n=1}^\infty
 \frac{\Gamma(D\!-\!3+n)}{\Gamma(\frac{D}{2}-1+n)}\frac{(1-y^{db}/4)^n}{n}
\, ,
\label{IA1:D}
\end{eqnarray}
and the $D \to 4$ limit of this result is
\begin{equation}
I[A^{db}_1]=-\frac{1}{8\pi^2}{\rm ln}\bigg(\frac{y^{db}}{4}\bigg)
\, .
\label{intA1D4}
\end{equation}
Since in the same limit
\begin{equation}
A^{ab}_1(y)=-\frac{1}{8\pi^2}\frac{1}{y^{ab}}
\, ,
\label{A1D4}
\end{equation}
the corresponding expression for $s_{\xi}$~(\ref{source sxi}) in $D=4$ is
\begin{equation}
s^{db}_{\xi}(x^{\prime\prime};x^\prime)
=\frac{1}{8\pi^2}\bigg\{(1\!-\!\xi)\bigg(\frac{2}{y^{db}}-1\bigg)
   -\big(3\!-\!\xi\big)
           {\rm ln}\bigg(\frac{y^{db}}{4}\bigg)\bigg\}(x^{\prime\prime};x^{\prime})
\, .
\label{sxiD4}
\end{equation}
Together with the choice for the Green's function for $A_2$,
which we take to be the de Sitter invariant 
one~(\ref{total solution GIA_2}--\ref{secondDis4}),
expressions~(\ref{A2:final}), (\ref{source sxi}) 
and~(\ref{sxiD4}) constitute the necessary ingredients
for the sought for de Sitter invariant photon
propagator~(\ref{propagator: tensor structures a}) in covariant gauges.
In this work we have chosen to construct the photon propagator as a convolution
in position space~(\ref{A2:final}), such that -- in Minkowski space --
it yields a standard algebraic form in momentum space. This is to be contrasted 
with the work of Allen and Jacobson~\cite{Allen:1986}
and Kahya and Woodard~\cite{Kahya:2005kj}, which represent $A_2$ as
a function of $y(x;x')$. From Eq.~(\ref{eom:A2:I}) one sees that 
$A_2(y)$ can be easily written as a double indefinite integral over $y$.
A discussion of the difficulties one faces when attempting to implement such 
a procedure (in covariant gauges and in the case of $D=4$) is presented in 
Appendix~\ref{An alternative method to calculate the scalar structure function A_2}.


\section{Response to a static point charge}
\label{Response to a static point charge}

Here we shall use the photon propagator~(\ref{propagator: tensor structures a}),
(\ref{total solution GIA_2}), (\ref{A2:final}), (\ref{IA1:D}) and~(\ref{source sxi})
to calculate the response to a static point charge
on de Sitter space in $D=4$. The relevant equation is the suitable generalization
of Eq.~(\ref{response on point chargea}) to curved space-times,
\begin{equation}
A_{\mu}(x)
=\int d^4 x' \sqrt{-g(x')} \big[{}_{\mu}\Delta_{\nu}^{\rm ret}\big](x;x')J^{\nu}(x')
\, ,
\label{potential}
\end{equation}
where the charge current density is given by
\begin{eqnarray}
J^{\nu}(x')=\frac{e\delta^{\nu}_0}{a'}
           \frac{\delta^3(\vec{x}^{\,\prime})}{\sqrt{\gamma(x')}}
\, .
\label{source dS}
\end{eqnarray}
Here $\gamma(x')=(a')^6$ is the determinant of the spatial part of the metric.
The retarded propagator needed for this is given by
\begin{equation}
\imath [{}_{\nu}\Delta^{\rm ret}_{\alpha}](x;x')
  =\imath  [{}_{\nu}\Delta^{++}_{\alpha}](x;x')-\imath [{}_{\nu}\Delta^{+-}_{\alpha}](x;x')\, ,
\end{equation}
where
\begin{eqnarray}
\label{photprop++}
  \imath [_{\nu}\Delta_{\alpha}^{ab}](x;x^\prime)
    &=& (\partial_{\nu}\partial^\prime_{\alpha} y)\times A_1^{ab}\big(y(x;x^\prime)\big)
    + \partial_{\nu}\partial^\prime_{\alpha} A_2^{ab}\big(y(x;x^\prime)\big)
\,,
\label{propagator: tensor structures}
\end{eqnarray}
and  $A_2^{ab}(x;x')$ is given in~(\ref{A2:final}).
Making use of
\begin{equation}
G_{A_2}^{\rm ret}=G_{A_2}^{++}-G_{A_2}^{+-}=G_{A_2}^{-+}-G_{A_2}^{--}
 \,;\qquad
s_{\xi}^{\rm ret}=s_{\xi}^{++}-s_{\xi}^{+-}=s_{\xi}^{-+}-s_{\xi}^{--}
\,,
\label{ret G and source}
\end{equation}
one easily finds
\begin{equation}
\imath [{}_{\nu}\Delta^{\rm ret}_{\alpha}](x;x')
  =(\partial_{\nu}\partial^\prime_{\alpha} y)\times A_1^{\rm ret}\big(y(x;x^\prime)\big)
    + \partial_{\nu}\partial^\prime_{\alpha} A_2^{\rm ret}\big(y(x;x^\prime)\big)
\, ,
\label{Delta ret}
\end{equation}
where
\begin{equation}
 A_1^{\rm ret}(x;x')= A_1^{++}(x;x')- A_1^{+-}(x;x')
 \,;\qquad
A_2^{\rm ret}(x;x')=H^2\int d^4x''G_{A_2}^{\rm ret}(x;x'')s_{\xi}^{\rm ret}(x'';x')
\,.
\label{ret G and source}
\end{equation}
Now from
\begin{eqnarray}
\frac{1}{y^{++}(x;x')}-\frac{1}{y^{+-}(x;x')}&=&\frac{1}{y^{-+}(x;x')}-\frac{1}{y^{--}(x;x')}
=-\frac{2\pi \imath \theta(\eta\!-\!\eta')}{H^2aa'}\delta(\|\vec x\!-\!\vec x^{\,\prime}\|^2-(\eta\!-\!\eta')^2)
\nonumber\\
\frac{1}{4-y^{++}(x;x')}-\frac{1}{4-y^{+-}(x;x')}
   &=&\frac{2\pi \imath \theta(\eta\!-\!\eta')}{H^2aa'}\delta\Big(\frac{4}{H^2aa'}-\|\vec x\!-\!\vec x^{\,\prime}\|^2+(\eta-\eta')^2\Big)
   \nonumber\\
   &=&\frac{\pi \imath \theta(\eta\!-\!\eta')}{H^2aa'\|\vec x\!-\!\vec x^{\,\prime}\|}
            \delta\Big(\eta\!+\!\eta'+\|\vec x\!-\!\vec x^{\,\prime}\|\Big)
\label{getting retarded}
\end{eqnarray}
and 
\begin{eqnarray}
{\rm ln}\bigg(\frac{y^{++}(x;x')}{y^{+-}(x;x')}\bigg)
&=&{\rm ln}\bigg(\frac{y^{-+}(x;x')}{y^{--}(x;x')}\bigg)
=2\pi \imath\theta(\eta\!-\!\eta') \theta\big(\eta\!-\!\eta'-\|\vec x\!-\!\vec x^{\,\prime}\|\big)
\label{getting retardedC:2}\\
{\rm ln}\frac{4-y^{++}(x;x')}{4-y^{+-}(x;x')}
  &=&-2\pi \imath \theta(\eta\!-\!\eta')\theta\Big(\|\vec x\!-\!\vec x^{\,\prime}\|^2-\frac{4}{H^2aa'}-(\eta\!-\!\eta')^2\Big)
   = -2\pi \imath \theta(\eta\!-\!\eta')\theta\big(\eta\!+\!\eta'+\|\vec x\!-\!\vec x^{\,\prime}\|\big)
\nonumber
\end{eqnarray}
and Eqs.~(\ref{secondDis4}), (\ref{A1D4}) and~(\ref{sxiD4}) we get
\begin{eqnarray}
A_1^{\rm ret}(x;x') &=& \frac{\imath }{8\pi}
         \frac{\theta(\eta\!-\!\eta')}{H^2aa'\|\vec x\!-\!\vec x^{\,\prime}\|}\delta(\eta\!-\!\eta'-\|\vec x\!-\!\vec x^{\,\prime}\|)
\label{A1:ret}
\\
G_{A_2}^{\rm ret}(x;x'') &=&-\frac{\theta(\eta\!-\!\eta'')}{4\pi}
 \Bigg(
     \frac{1}{aa''\|\vec x\!-\!\vec x^{\,\prime\prime}\|}
       \Big(\delta(\eta\!-\!\eta''-\|\vec x\!-\!\vec x^{\,\prime\prime}\|)
            +\delta\big(\eta\!+\!\eta''+\|\vec x\!-\!\vec x^{\,\prime\prime}\|\big)\Big)
            \nonumber\\
&&\hskip 1.8cm
     +\,H^2 \theta\big(\eta\!-\!\eta''-\|\vec x\!-\!\vec x^{\,\prime\prime}\|\big)
      + H^2\theta\big(\eta\!+\!\eta''+\|\vec x\!-\!\vec x^{\,\prime\prime}\|\big)
 \Bigg)
\label{GA2:ret}\\
s_\xi^{\rm ret}(x'';x') &=& -\imath \frac{\theta(\eta''\!-\!\eta')}{4\pi H^2}
\bigg\{\frac{1\!-\!\xi}{aa''\|\vec x^{\,\prime\prime}\!-\!\vec x^{\,\prime}\|}
       \delta(\eta''\!-\!\eta'-\|\vec x^{\,\prime\prime}\!-\!\vec x^{\,\prime}\|)
+\big(3\!-\!\xi\big)H^2\theta\big(\eta''\!-\!\eta'-\|\vec x^{\,\prime\prime}\!-\!\vec x^{\,\prime}\|\big)
\bigg\}
\,.\qquad
\label{sxi:ret}
\end{eqnarray}
Upon inserting these expressions into
Eq.~(\ref{propagator: tensor structures}) and (\ref{A2:final}) one gets, 
\begin{eqnarray}
 [{}_{\nu}\Delta^{\rm ret}_{\alpha}](x;x^{\prime})
&=&\frac{\partial_\nu\partial_\alpha^\prime y(x;x')}{2H^2aa'}
\frac{\theta\big(\eta\!-\!\eta'\big)}{4\pi \|\vec x\!-\!\vec x^{\,\prime}\|}
    \delta\big(\eta\!-\!\eta'-\|\vec x\!-\!\vec x^{\,\prime}\|\big)
\label{propagator:ret final}
 \\
&&
+\,\frac{1}{(4\pi)^2}\partial_{\nu}\partial'_{\alpha}\int d^4 x^{\,\prime\prime} \Bigg\{
\theta\big(\eta\!-\!\eta''\big)
\Bigg[\frac{1}{aa''\|\vec x\!-\!\vec x^{\,\prime\prime}\|}
\Big(\delta\big(\eta\!-\!\eta''\!-\!\|\vec x\!-\!\vec x^{\,\prime\prime}\|\big)
+\delta\big(\eta\!+\!\eta''\!+\!\|\vec x\!-\!\vec x^{\,\prime\prime}\|\big )
\Big)
\nonumber \\
&&\hskip 5.cm
+\,H^2\theta\big(\eta\!-\!\eta''-\|\vec x\!-\!\vec x^{\,\prime\prime}\|\big)
+H^2\theta\big(\eta\!+\!\eta''+\|\vec x\!-\!\vec x^{\,\prime\prime}\|\big)
\Bigg]
\nonumber \\
&&\hskip 1.8cm
\times\,
\theta\big(\eta''\!-\!\eta'\big)\Bigg[\frac{1\!-\!\xi}{a'a''\|\vec x^{\,\prime\prime}\!-\!\vec x^{\,\prime}\|}
\delta\big(\eta''\!-\!\eta'-\|\vec x^{\,\prime\prime}\!-\!\vec x^{\,\prime}\|\big)
+(3\!-\!\xi)H^2\theta\big(\eta''\!-\!\eta'\!
                 -\!\|\vec x^{\,\prime\prime}\!-\!\vec x^{\,\prime}\|\big)
\Bigg]
\Bigg\}
\nonumber\\
&\equiv& {\cal \aleph}_{\nu\alpha}(x;x') 
            + \partial_\nu\partial_\alpha^\prime\lambda(x;x')
\,.
\nonumber
\end{eqnarray}
From the form of this expression we see that, 
just as in the Minkowski case~(\ref{retarder propagator:Minkowski:3}), 
only the (gauge independent) term in the first line 
can contribute to physical quantities, while the other terms
which depend on $\xi$ cannot contribute. 

To illustrate how this works in some detail we shall now calculate
the electromagnetic response to a static point charge~(\ref{source dS}) on 
four dimensional de Sitter space.  
From~(\ref{potential}) and~(\ref{propagator:ret final}) we see that 
the gauge independent contribution to the electromagnetic potential 
\begin{eqnarray}
\tilde{\cal A}_\mu(x) &=& \int d^4x' \sqrt{-g(x')}{\cal \aleph}_{\nu\alpha}(x;x')J^\alpha(x')
\label{potential dS:gauge independent}
\\
&=& \frac{e}{8\pi\|\vec{x}\|}\int_{\eta_o}^{\eta} d\eta'
\bigg\{\Big[y(x;x')\delta_{\mu}^0+2\delta_{\mu}^0a(\eta)H(\eta\!-\!\eta')
+2a(\eta')H\eta_{\mu\alpha}(x^{\alpha}\!-\!x'^{\alpha})-2\eta_{\mu 0}\Big]
     \theta(\eta\!-\!\eta')\delta\big(\eta\!-\!\eta'\!-\!\|\vec{x}\|\big)\bigg\}\, .
\nonumber
\end{eqnarray}
where we integrated over $\vec x^{\,\prime}$.  
If we introduce $r=\|\vec{x}\|$ and $\Delta\eta=\eta-\eta'$
and change the variable of integration to $\Delta \eta$, 
Eq.~(\ref{potential dS:gauge independent}) becomes, 
\begin{eqnarray}
\tilde{\cal A}_{\mu}&=&\frac{e}{8\pi r}\int_{0}^{\Delta\eta_0} d(\Delta\eta)
\bigg\{\Big[y(x;x')\delta_{\mu}^0+2\delta_{\mu}^0a(\eta)H\Delta\eta
+2a(\eta\!-\!\Delta\eta)H\eta_{\mu\alpha}(x^{\alpha}\!-\!x'^{\alpha})
-2\eta_{\mu 0}\Big]\delta\big(\Delta\eta\!-\!r\big)\bigg\}
\quad
\, ,
\label{potential dS:gauge independent:2}
\end{eqnarray}
where $\Delta\eta_0=\eta-\eta_0$. Since $\Delta\eta_0 \geq \Delta\eta \geq 0$,
Eq.~(\ref{potential dS:gauge independent:2}) is easily integrated to yield,
\begin{eqnarray}
\tilde{\cal A}_{\mu}&=&\frac{e\theta(\Delta\eta_0\!-\!r)}{8\pi r}
\Big[2\delta_{\mu}^0a(\eta)H\Delta\eta
+2a(\eta\!-\!\Delta\eta)H\eta_{\mu\alpha}(x^{\alpha}\!-\!x'^{\alpha})
-2\eta_{\mu 0}\Big]_{\Delta\eta=r} 
\label{potential dS:gauge independent:3}
\end{eqnarray}
For the time-like component we get
\begin{eqnarray}
\tilde{\cal A}_{0}&=&\frac{e}{4\pi}
\bigg(\frac{1}{r}-\frac{1}{\eta}+\frac{1}{\eta-r}\bigg)\theta(\Delta\eta_0\!-\!r)
\, ,
\label{pot0}
\end{eqnarray}
and for the space-like components we obtain
\begin{eqnarray}
\tilde{\cal A}_{i}&=&- \frac{e}{4\pi }\frac{x^i}{r(\eta-r)}\theta(\Delta\eta_0-r)
\, .
\label{poti}
\end{eqnarray}
As in the Minkowski case, the physical result is recovered when
$\eta_0\rightarrow -\infty$,
in which case~(\ref{pot0}--\ref{poti}) reduce to, 
\begin{equation}
\tilde{\cal A}_{0}=\frac{e}{4\pi}
\bigg(\frac{1}{r}-\frac{1}{\eta}+\frac{1}{\eta\!-\!r}\bigg)
\, ,\qquad
\tilde{\cal A}_{i}=- \frac{e}{4\pi }\frac{x^i}{r(\eta\!-\! r)}
\, .
\label{pot:B}
\end{equation}
A part of this potential is pure gauge. Indeed,
Eqs.~(\ref{pot:B}) can be recast as 
\begin{equation}
\tilde{\cal A}_{0}= {\cal A}_{0}+\partial_0\Lambda_1(x)
\, , \qquad
\tilde{\cal A}_{i}=\partial_i \Lambda_1(x)
\, , \qquad
{\cal A}_{0}=\frac{e}{4\pi r}
\, , \qquad
\Lambda_1(x)=\frac{e}{4\pi}\ln \bigg(1-\frac{r}{\eta}\bigg)
\, .  
\label{pot:C}
\end{equation}
such that the gauge independent part of the potential on de Sitter is simply,
\begin{equation}
 {\cal A}_\mu(x) =  \frac{e}{4\pi r}\delta_\mu^{\;0}
\, . 
\label{pot:D}
\end{equation}
The second part of the response potential arises from the gauge dependent 
part of the propagator~(\ref{propagator:ret final}), and it is of the form,
\begin{eqnarray}
\delta A_\mu(x) 
= e\partial_\mu \int d\eta^\prime \partial_0^\prime \lambda(x;x')
  |_{\vec x^{\,\prime}\rightarrow 0}
\equiv \partial_\mu \Lambda_2(x)
\,,
\label{dS potential:second part}
\end{eqnarray}
and hence it is pure gauge. This conclusion is legitimate, provided
the $\eta^\prime$ and $x''$ integrals (the latter 
appearing in~(\ref{propagator:ret final}))
are all finite, which we have checked to be the case. 

To summarize, we have found that the response to a point charge on de Sitter
(present from $\eta_0\rightarrow -\infty$)
yields an electric field that is conformal to the Minkowski response, 
and a vanishing magnetic field, 
\begin{equation}
A_\mu(x) = \frac{e}{4\pi r}\delta_\mu^{\;0} + \partial_\mu \Lambda(x)
\,,\qquad 
E_i = F_{0i} = \partial_0 A_i - \partial_i A_0 = \frac{e}{4\pi r^2}\frac{r^i}{r}
\,,\qquad 
B_{i} = \epsilon_{ijl}\partial_j A_l = 0
\,,
\label{dS:E and B}
\end{equation}
where $\Lambda=\Lambda_1+\Lambda_2$.
From these results we see that, because electromagnetism is conformal on 
cosmological spaces in $D=4$, (apart from the conformal rescaling 
of the fields by a power 
of the scale factor) the expansion of the Universe plays no role in 
the electromagnetic fields generated by a static point electric charge.


\section{The one-loop stress energy tensor}
\label{One loop Tmn}

 In this section we calculate the one-loop stress energy tensor
from our propagator. This is in principle
important for the calculation of the quantum backreaction on de Sitter space,
albeit we do not expect a large backreaction from photons, since
they couple conformally in four space-time dimensions.

 We start with the well known formula for the photon stress energy
tensor,
\begin{equation}
T_{\mu\nu} = \Big(\delta_\mu^{\;\alpha}\delta_\nu^{\;\gamma}g^{\beta\delta}
             - \frac14 g_{\mu\nu}g^{\alpha\gamma}g^{\beta\delta}
            \Big)F_{\alpha\beta}F_{\gamma\delta}
\,,
\label{photon stress energy}
\end{equation}
which relates $T_{\mu\nu}$ to $F_{\alpha\beta}F_{\gamma\delta}$.
The one-loop contribution to the expectation value
for this quadratic operator can be obtained from,
\begin{equation}
 \langle\Omega|F^a_{\alpha\beta}(x)F^b_{\gamma\delta}(x)
          |\Omega\rangle_{\rm 1 \; loop}
=\Big\{\partial^\prime_\gamma
     \Big(\partial_\alpha[\imath{}_\beta\Delta^{ab}_\delta](x;x^\prime)
         -\partial_\beta[\imath{}_\alpha\Delta^{ab}_\delta](x;x^\prime)\Big)
-\partial^\prime_\delta
     \Big(\partial_\alpha[\imath{}_\beta\Delta^{ab}_\gamma](x;x^\prime)
         -\partial_\beta[\imath{}_\alpha\Delta^{ab}_\gamma](x;x^\prime)\Big)
\Big\}_{x^\prime\rightarrow x}
\,.
\label{photon stress energy:2}
\end{equation}
Taking account of Eqs.~(\ref{propagator: tensor structures}),
(\ref{A1:photon solution}), (\ref{photon Keldysh propagator:A1}) 
and~(\ref{A1:photon solution:light-cone}), 
we see from
\begin{equation}
 \imath[{}_\beta\Delta^{ab}_\delta](x;x^\prime)
         = (\partial_\beta\partial^\prime_\delta y) A_1(y^{ab})
         + \partial_\beta\partial^\prime_\delta A_2(y^{ab})
\label{A1:photon solution:new}
\end{equation}
that, because of the antisymmetrization of the indices $\alpha$ and $\beta$,
the $A_2$-term does not contribute to the expectation value
in~(\ref{photon stress energy:2}). Furthermore, since
$F_{\mu\nu}=\partial_\mu A_\nu - \partial_\nu A_\mu
=\nabla_\mu A_\nu - \nabla_\nu A_\mu$, all of the ordinary derivatives
in~(\ref{photon stress energy:2}) can be replaced by covariant derivatives.
With these remarks, Eq.~(\ref{photon stress energy:2}) becomes
\begin{eqnarray}
 \langle\Omega|F^a_{\alpha\beta}(x)F^b_{\gamma\delta}(x)
  |\Omega\rangle_{\rm 1 \; loop}
&=&\Big\{\nabla^\prime_\gamma
     \Big(\nabla_\alpha[(\partial_\beta\partial^\prime_\delta y) A_1(y^{ab})]
     -\nabla_\beta[(\partial_\alpha\partial^\prime_\delta y) A_1(y^{ab})]\Big)
\nonumber\\
&&\hskip 0cm
-\,\nabla^\prime_\delta
     \Big(\nabla_\alpha[(\partial_\beta\partial^\prime_\gamma y) A_1(y^{ab})]
         -\nabla_\beta[(\partial_\alpha\partial^\prime_\gamma y) A_1(y^{ab})]
\Big)
\Big\}(x;x^\prime)_{x^\prime\rightarrow x}
\,.
\label{photon stress energy:3}
\end{eqnarray}
Any dependence on the gauge parameter $\xi$
(which is in the structure function $A_2$) has dropped out,
such that~(\ref{photon stress energy:3}) is manifestly gauge independent.
Now, since $\nabla_\alpha\partial_\beta\partial^\prime_\delta y
   = -H^2g_{\alpha\beta}(x)\partial^\prime_\delta y $ is symmetric in
$\{\alpha,\beta\}$, the first covariant
derivatives in~(\ref{photon stress energy:3}) act only on $A_1$,
resulting in,
\begin{eqnarray}
 \langle\Omega|F^a_{\alpha\beta}(x)F^b_{\gamma\delta}(x)
  |\Omega\rangle_{\rm 1 \; loop}
&=&\Big\{\nabla^\prime_\gamma
     \Big((\partial_\beta\partial^\prime_\delta y)(\partial_\alpha y)
              A_1^\prime(y^{ab})
     -(\partial_\alpha\partial^\prime_\delta y)(\partial_\beta y)
              A_1^\prime(y^{ab})\Big)
\nonumber\\
&&\hskip 0cm
-\,\nabla^\prime_\delta
     \Big((\partial_\beta\partial^\prime_\gamma y)(\partial_\alpha y)
                A_1^\prime(y^{ab})
         -(\partial_\alpha\partial^\prime_\gamma y)(\partial_\beta y)
                    A_1^\prime(y^{ab})
\Big)
\Big\}_{x^\prime\rightarrow x}
\,.
\label{photon stress energy:4}
\end{eqnarray}
Analogously, because of the symmetry in the primed indices,
the primed covariant derivatives in~(\ref{photon stress energy:4})
commute through the first terms and one gets,
\begin{eqnarray}
 \langle\Omega|F^a_{\alpha\beta}(x)F^b_{\gamma\delta}(x)
  |\Omega\rangle_{\rm 1 \; loop}
&=&\Big\{
4(\partial_\alpha\partial^\prime_{[\gamma}y)
           (\partial^\prime_{\delta]}\partial_\beta y)A_1^\prime(y^{ab})
- 4(\partial_{[\alpha} y)(\partial_{\beta]}\partial^\prime_{[\gamma} y)
                  (\partial^\prime_{\delta]} y)A_1^{\prime\prime}(y^{ab})
\Big\}_{x^\prime\rightarrow x}
\label{photon stress energy:5}
\end{eqnarray}
Because of the equalities,
\begin{equation}
 (\partial^\prime_{\delta}\partial_\beta y)_{x^\prime\rightarrow x}
    = -2H^2g_{\delta\beta}(x)
\,,\qquad
 (\partial_\beta y)_{x^\prime\rightarrow x} =0
\,,
\label{coincident derivatives of y}
\end{equation}
the second term in~(\ref{photon stress energy:5}) vanishes
and we find,
\begin{equation}
 \langle\Omega|F^a_{\alpha\beta}(x)F^b_{\gamma\delta}(x)
  |\Omega\rangle_{\rm 1 \; loop}
= 16H^4(g_{\alpha[\gamma}g_{\delta]\beta})
              [A_1^\prime(y^{ab})]_{x^\prime\rightarrow x}
\,.
\label{photon stress energy:6}
\end{equation}
From Eq.~(\ref{A1:photon solution:light-cone}) we see that
the coincident limit of $A_1^\prime(y)$ equals,
\begin{equation}
 [A_1^\prime(y^{ab})]_{x^\prime\rightarrow x}
    = \frac{1}{8}\frac{H^{D-4}}{(4\pi)^{D/2}}
    \frac{\Gamma(D\!-\!1)}{\Gamma\big(\frac{D}{2}\!+\!1\big)}
\;\;\stackrel{D\rightarrow 4}{\longrightarrow}\;\;
\frac{1}{128\pi^2}
\,,
\label{coincident limit of A1}
\end{equation}
where, to obtain this result in the spirit of dimensional regularization we
have assumed that $\Re[D]<0$ (recall that this procedure of analytic
continuation subtracts automatically all power law divergences).
When Eq.~(\ref{coincident limit of A1}) is inserted 
into~(\ref{photon stress energy:6}), one obtains
\begin{equation}
 \langle\Omega|F^a_{\alpha\beta}(x)F^b_{\gamma\delta}(x)
  |\Omega\rangle_{\rm 1 \; loop}
=\frac{2H^{D}}{(4\pi)^{D/2}}\frac{\Gamma(D\!-\!1)}
                                 {\Gamma\big(\frac{D}{2}\!+\!1\big)}
                                  \times g_{\alpha[\gamma}g_{\delta]\beta}(x)
\;\;\stackrel{D\rightarrow 4}{\longrightarrow}\;\;
\frac{H^4}{8\pi^2}\times g_{\alpha[\gamma}g_{\delta]\beta}(x)
\,,
\label{photon stress energy:7}
\end{equation}
independent on the polarities $a,b=\pm$, as was to be expected for
a one-loop coincident correlator. Finally, upon
inserting~(\ref{photon stress energy:7}) into~(\ref{photon stress energy}),
one gets for the renormalized one-loop stress energy tensor,
\begin{equation}
\langle\Omega|T_{\mu\nu}(x)|\Omega\rangle_{\rm 1\;loop}
=-\frac{H^{D}}{(4\pi)^{D/2}}\frac{\Gamma(D)}
                                 {\Gamma\big(\frac{D}{2}\!+\!1\big)}
                                 \frac{D\!-\!4}{4} g_{\mu\nu}(x)
\;\;\stackrel{D\rightarrow 4}{\longrightarrow}\;\;
0
\,.
\label{photon stress energy:8}
\end{equation}
This means that, at the one-loop order and in $D=4$, the electromagnetic
stress-energy tensor on de Sitter space does not break conformal symmetry.
Namely, when taken together, the de Sitter and conformal symmetry
require that the expectation value of $T_{\mu\nu}$
taken in a de Sitter invariant state
must vanish. That it must be so can be argued as follows.
The de Sitter symmetry alone requires
$\langle\Omega|T_{\mu\nu}(x)|\Omega\rangle = T(x) g_{\mu\nu}$, where
$T$ is some scalar function of $x$. But the only scalar function of $x$
consistent with the de Sitter symmetry is a constant, $T(x)=T_0$.
Next, the conformality of $\langle\Omega|T_{\mu\nu}(x)|\Omega\rangle$
requires that its trace must vanish,
$g^{\mu\nu}\langle\Omega|T_{\mu\nu}(x)|\Omega\rangle = 4T_0=0$, from which
it immediately follows that $T_0=0$, 
in agreement with~(\ref{photon stress energy:8}).
Of course, the results~(\ref{photon stress energy:7}--\ref{photon stress energy:8}) 
are not new; compare, for example,
Eqs.~(\ref{photon stress energy:7}--\ref{photon stress energy:8})
with Eqs.~(40) and (42) in Ref.~\cite{Prokopec:2008gw}.
The new ingredient here is the photon propagator
in general covariant gauges, which we used to show that
the results~(\ref{photon stress energy:7}--\ref{photon stress energy:8}) are
fully gauge independent.


\section{Conclusions and discussion}
\label{Conclusions and discussion}

In this paper we have found a de Sitter invariant propagator
in general covariant gauges. We have used it to calculate the response to 
a static point charge, and the one loop stress energy tensor. 
Also the gauge independence of the results is demonstrated. 
In searching for the de Sitter invariant solution we needed to introduce
an unphysical source at the antipodal point in one of the steps. 
As it turned out this had no bearing on physical results since 
the contribution of those terms appears only in the pure gauge part
of the propagator. Our propagator sheds some light on the question
whether it is possible to find de Sitter invariant propagators in gauge theories
and allows for perturbative calculations on de Sitter, which is 
the model space for inflation. 
Although the results presented here are interesting in their own right, 
what we really hope for is that an analogous treatment can be applied 
to construct a graviton propagator on de Sitter in general covariant gauges.
Indeed, constructing a photon propagator in covariant gauges is a first step 
towards constructing a graviton propagator in general covariant gauges.
Apart from the difference in the number of vector indices and de Sitter 
invariant tensors, the photon and graviton propagators differ in one more 
important aspect. Namely, while in de Sitter space and near four space-time
dimensions electromagnetism couples to de Sitter in a nearly conformal manner,
such that the effect of Universe's expansion on the creation of photons 
is nearly minimal, gravitons strongly break conformal invariance, implying  
abundant production of gravitons on de Sitter background,  
strongly suggesting that de Sitter symmetry is broken
in (perturbative) quantum gravity.
Therefore, understanding the interplay between de Sitter breaking,
gauging, and graviton production is of a crucial importance to 
understanding the stability of de Sitter space. 
Recently graviton propagators were constructed in (generalized) de Donder
gauges~\cite{Mora:2012zi,Miao:2011fc} as well as in a 
non-covariant gauge~\cite{Tsamis:2005je}. These propagators can be used to 
study loop quantum effects on de Sitter space~\cite{Mora:2013ypa}, 
as well as the stability of de Sitter space. 
The problem that still remains elusive is a general proof
that de Sitter symmetry must be broken by gravitons.
If proved, this theorem would immediately imply a (perturbative) instability 
of de Sitter space. 
Already a long time ago Allen and Folacci~\cite{Allen:1987tz}
have shown that a minimally coupled massless
scalar field necessarily breaks de Sitter symmetry, 
thus perturbatively destabilizing de Sitter space. 
While the existence of minimally coupled scalar fields
can be questioned, it is very hard to argue against the existence of
gravitons, which underpins the importance of understanding
perturbative stability of quantum gravity on de Sitter space.


\appendix


\section{The photon propagator on Minkowski background}
\label{The photon propagator on Minkowski background}

In this appendix we perform explicit integrations of 
section~\ref{Introduction}. Notice first that Eq.~(\ref{response to point charge})
can be written in the following form, 
\begin{eqnarray}
 A_\mu(x)&=&\frac{e\delta_{\mu}^0}{4\pi r}\theta(\Delta\eta_0-r)
-\,(1\!-\!\xi)\frac{e}{8\pi}\partial_\mu\partial_0
 \int_{\eta_0}^\eta d\eta^{\prime\prime}
   \int_{\eta_0}^{\eta^{\prime\prime}} d\eta^{\prime}
        \int_{-1}^1dz
    \frac{\delta\left((\eta\!-\!\eta^{\prime\prime})
                          -\|\vec x\!-\!\vec x^{\,\prime\prime}\,\|\right)}
           {\|\vec x\!-\!\vec x^{\,\prime\prime}\,\|}
(\eta^{\prime\prime}\!-\!\eta^\prime)
\,,
\label{response to point charge:2}
\end{eqnarray}
where $z=\cos[\angle(\vec x, \vec x^{\,\prime\prime})]$,
$\|\vec x\!-\!\vec x^{\,\prime\prime}\,\|
  =\sqrt{r^2+r^{\prime\prime 2}-2rr^{\prime\prime}z}$,
and we have used the second delta-function in the second line of
 Eq.~(\ref{response to point charge}) 
to integrate over $r^{\prime\prime}$, whereby
$r^{\prime\prime}\rightarrow \eta^{\prime\prime}-\eta^\prime$.
Next, we shall use the delta-function in~(\ref{response to point charge:2})
to integrate over $z$. In doing so, first notice that
\begin{equation}
    \frac{\delta\left((\eta\!-\!\eta^{\prime\prime})
                          -\|\vec x\!-\!\vec x^{\,\prime\prime}\,\|\right)}
           {\|\vec x\!-\!\vec x^{\,\prime\prime}\,\|}
 = \frac{\delta(z-z_+)+\delta(z-z_-)}{r(\eta^{\prime\prime}-\eta^\prime)}
\,,
\label{response to point charge:3}
\end{equation}
where $z=z_{\pm}$ are the two poles of at which the argument of the
delta-function vanishes. Notice that the delta-function gives a
contribution to the integral only if the poles lie in the interval of
integration,
\begin{equation}
 -1<z_\pm<1
\,,
\label{constraint on z+-}
\end{equation}
which can be also written as,
\begin{equation}
 r^2-2r(\eta^{\prime\prime}\!-\!\eta^\prime)
 + (\eta^{\prime\prime}\!-\!\eta^\prime)^2 - (\eta\!-\!\eta^{\prime\prime})^2
  < 0
<  r^2+2r(\eta^{\prime\prime}\!-\!\eta^\prime)
 + (\eta^{\prime\prime}\!-\!\eta^\prime)^2 - (\eta\!-\!\eta^{\prime\prime})^2
\,,
\label{constraint on z+-:2}
\end{equation}
The left inequality is satisfied for $r_-<r<r_+$, where
\begin{equation}
 r_-=(\eta^{\prime\prime}\!-\!\eta^\prime) - (\eta\!-\!\eta^{\prime\prime})
\,,\qquad
 r_+=(\eta^{\prime\prime}\!-\!\eta^\prime) + (\eta\!-\!\eta^{\prime\prime})
    = \eta\!-\!\eta^\prime
\,,
\label{constraint on z+-:3}
\end{equation}
while the right inequality in~(\ref{constraint on z+-:2}) is satisfied when
$r>\tilde r_+$ or when $r<\tilde r_-$, where
\begin{equation}
\tilde r_-
 =-(\eta^{\prime\prime}\!-\!\eta^\prime) - (\eta\!-\!\eta^{\prime\prime})
  = -(\eta\!-\!\eta^\prime)
\,,\qquad
\tilde r_+
  =-(\eta^{\prime\prime}\!-\!\eta^\prime) + (\eta\!-\!\eta^{\prime\prime})
\,,
\label{constraint on z+-:4}
\end{equation}
and of course $r>0$ (notice that both $z_+$ and $z_-$ result in
the same constraints on $r$). This means that
\begin{equation}
-(\eta^{\prime\prime}\!-\!\eta^\prime) + (\eta\!-\!\eta^{\prime\prime})
  < r < \eta\!-\!\eta^\prime
\,,
\label{constraint on z+-:5}
\end{equation}
and of course $r>0$. These conditions can be written as
\begin{equation}
 \theta\big(-(\eta^{\prime\prime}\!-\!\eta^\prime)
            +(\eta\!-\!\eta^{\prime\prime})\big)
   \big[\theta(r+(\eta^{\prime\prime}\!-\!\eta^\prime)
            -(\eta\!-\!\eta^{\prime\prime}))
    - \theta\big(r-(\eta\!-\!\eta^\prime)\big)\big]
 +\theta\big((\eta^{\prime\prime}\!-\!\eta^\prime)
            -(\eta\!-\!\eta^{\prime\prime})\big)
   \big[\theta(\eta\!-\!\eta^\prime-r)\big]
\label{constraint on z+-:6}
\end{equation}
With these remarks in mind, we can perform the
$z$-integral in Eq.~(\ref{response to point charge:2}),
\begin{eqnarray}
 A_\mu(x)&=&\frac{e\delta_{\mu}^0}{4\pi r}\theta(\Delta\eta_0-r)
-(1\!-\!\xi)\frac{e}{4\pi}\partial_\mu\partial_0 \frac{1}{r}
 \int_{\eta_0}^\eta d\eta^{\prime\prime}
   \int_{\eta_0}^{\eta^{\prime\prime}} d\eta^{\prime}
\label{response to point charge:4}
\\
&&\hskip -1.5cm
 \times\,\Big\{\theta\big(\!-\!(\eta^{\prime\prime}\!-\!\eta^\prime)
            +(\eta\!-\!\eta^{\prime\prime})\big)
   \big[\theta(r+(\eta^{\prime\prime}\!-\!\eta^\prime)
            -(\eta\!-\!\eta^{\prime\prime}))
    - \theta\big(r\!-\!(\eta\!-\!\eta^\prime)\big)\big]
 +\theta\big((\eta^{\prime\prime}\!-\!\eta^\prime)
            -(\eta\!-\!\eta^{\prime\prime})\big)
   \big[\theta(\eta\!-\!\eta^\prime\!-\!r)\big]
\Big\}
\,.
\nonumber
\end{eqnarray}
We can now act with the time derivative to get rid of one of the integrals,
\begin{eqnarray}
 A_\mu(x)&=&\frac{e\delta_{\mu}^0}{4\pi r}\theta(\Delta\eta_0-r)
-(1\!-\!\xi)\frac{e}{4\pi}\partial_\mu
   \int_{\eta_0}^\eta \frac{d\eta^{\prime}}{r}
 \Big\{\theta(\eta\!-\!\eta^\prime\!-\!r)
\Big\}
\,.
\nonumber\\
 &&-\,(1\!-\!\xi)\frac{e}{4\pi}\partial_\mu \frac{1}{r}
 \int_{\eta_0}^\eta d\eta^{\prime\prime}
   \int_{\eta_0}^{\eta^{\prime\prime}}\! d\eta^{\prime}
 \Big\{\theta\big(\!-\!(\eta^{\prime\prime}\!-\!\eta^\prime)
            \!+\!(\eta\!-\!\eta^{\prime\prime})\big)
   \big[\!-\!\delta(r\!+\!(\eta^{\prime\prime}\!-\!\eta^\prime)
            \!-\!(\eta\!-\!\eta^{\prime\prime}))
       +\delta(r\!-\!(\eta\!-\!\eta^\prime))\big]
\nonumber\\
&&\hskip 5cm
 +\,\theta\big((\eta^{\prime\prime}\!-\!\eta^\prime)
            \!-\!(\eta\!-\!\eta^{\prime\prime})\big)
      \delta((\eta\!-\!\eta^\prime)\!-\!r)
\Big\}
\,.
\label{response to point charge:5}
\end{eqnarray}
The $\eta^\prime$-integral in the last two lines can be performed,
and the result is
\begin{eqnarray}
 &&\hskip -0.3cm
-\,(1\!-\!\xi)\frac{e}{4\pi}\partial_\mu \frac{1}{r}
 \int_{\eta_0}^\eta d\eta^{\prime\prime}
   \int_{\eta_0}^{\eta^{\prime\prime}} d\eta^{\prime}
 \Big\{\!-\!\delta(r\!+\!(\eta^{\prime\prime}\!-\!\eta^\prime)
            \!-\!(\eta\!-\!\eta^{\prime\prime}))
       +\delta(r\!-\!(\eta\!-\!\eta^\prime))
 \Big\}
\label{response to point charge:5b}
\\
 &&\hskip 2.cm
=\,-(1\!-\!\xi)\frac{e}{4\pi}\partial_\mu \frac{1}{r}
 \int_{\eta_0}^\eta d\eta^{\prime\prime}
 \Big\{\!-\!\theta\big(r\big) 
 \theta \big(r-(\eta\!-\!\eta'')+(\eta''\!-\!\eta_0)\big)
+2\theta\big(r-(\eta\!-\!\eta'')\big)-\theta\big(r\!-\!(\eta\!-\!\eta_0)\big)\Big\}
\nonumber
\end{eqnarray}
\\
Collecting all the terms together we get,
\begin{eqnarray}
 A_\mu(x)
=\frac{e\delta_{\mu}^0}{4\pi r}\theta(\Delta\eta_0\!-\!r)
-(1\!-\!\xi)\frac{e}{4\pi}\partial_\mu\frac{1}{r}\!
   \int_{0}^{\Delta \eta_0}\!\! d\Delta\eta
 \Big\{\theta(\Delta\eta\!-\!r)
-         
\theta(r\!-\!2\Delta\eta\!+\!\Delta\eta_0)
   +2\theta(r\!-\!\Delta\eta)-\theta(r\!-\!\Delta\eta_0)
\Big\}
\,,\qquad
\end{eqnarray}
where $\Delta\eta = \eta-\eta^\prime$ and $\Delta\eta_0=\eta-\eta_0$m, $r\geq0$.
This result appears in the main text as (\ref{response to point charge:6}) and it is
then used to calculate the response potential $A_\mu(x)$.


\section{The photon tensor structures}
\label{The photon tensor structures}

Here we consider how the operator~(\ref{photoper})
on de Sitter space acts on the photon propagator
$\imath [_{\mu}\Delta_{\alpha}](x;x^\prime)$, where
the tensor decomposition (\ref{propagator: tensor structures a})
suitable for de Sitter space is used.

 The first operator (d'Alembertian) in~(\ref{photon propagator:eom})
acts as,
\begin{eqnarray}
&&  g^{\mu\nu}\Box_x\{(\partial_{\nu}\partial^\prime_{\alpha} y)\times f_1(y)
          + (\partial_{\nu}y)\times(\partial^\prime_{\alpha} y)\times f_2(y)\}
\label{Box on propagator}
\\
&&\hskip 4cm
= \, H^2(\partial^{\mu}\partial^{\prime}_{\alpha}y)
       \Big\{\big[(4y\!-\!y^2)f_1^{\prime\prime}
         +D(2\!-\!y)f_1^{\prime}-f_1\Big] + \Big[2(2\!-\!y)f_2\big]
    \Big\}
\nonumber\\
&&\hskip 4cm
= \, H^2(\partial^{\mu}y)\times(\partial^{\prime}_{\alpha}y)
 \Big\{\big[-2f_1'\big]
   +\big[(4y\!-\!y^2)f_2^{\prime\prime}+(D+4)(2-y)f_2^{\prime}-(D+1)f_2
\big]
    \Big\}
\,.
\nonumber
\end{eqnarray}
The {\it second part} of the operator in~(\ref{photoper}) acts as
\begin{eqnarray}
&&\nabla^{\nu}\nabla^{\mu}
   \{(\partial_{\nu}\partial^\prime_{\alpha} y)\times f_1(y)
      + (\partial_{\nu}y)\times(\partial^\prime_{\alpha} y)\times f_2(y)\}
\label{tensor structure:2}\\
    && \hskip 3cm
= \, H^2(\partial^{\mu}\partial^{\prime}_{\alpha}y)
    \Big\{\big[
         (2\!-\!y)f_1^{\prime}-f_1\big]
         +\big[(4y\!-\!y^2)f_2^{\prime}+(D+1)(2-y)f_2\big]\Big\}
\nonumber \\
&&\hskip 3cm
  +\,H^2(\partial^{\mu}y)(\partial^{\prime}_{\alpha}y)
    \Big\{\big[(2-y)f_1^{\prime\prime}-(D+1)f_1^{\prime}\big]
      +\big[(4y\!-\!y^2)f_2^{\prime\prime}+(D+3)(2-y)f_2^{\prime}-2f_2\big]
      \Big\}
\nonumber
\,.
\end{eqnarray}
Finally, the {\it third part} of the operator in~(\ref{photoper}) yields
\begin{eqnarray}
&&\nabla^\mu\nabla^\nu
   \{(\partial_{\nu}\partial^\prime_{\alpha} y)\times f_1(y)
      + (\partial_{\nu}y)\times(\partial^\prime_{\alpha} y)\times f_2(y)\}
\label{tensor structure:3}\\
    && \hskip 3cm
= \, H^2(\partial^{\mu}\partial^{\prime}_{\alpha}y)
    \Big\{\big[
         (2\!-\!y)f_1^{\prime}-Df_1\big]
         +\big[(4y\!-\!y^2)f_2^{\prime}+(D\!+\!1)(2\!-\!y)f_2\big]
\Big\}
\nonumber \\
&&\hskip 3cm
  +\,H^2(\partial^{\mu}y)(\partial^{\prime}_{\alpha}y)
    \Big\{\big[(2\!-\!y)f_1^{\prime\prime}-(D\!+\!1)f_1^{\prime}\big]
+\big[(4y\!-\!y^2)f_2^{\prime\prime}+(D\!+\!3)(2\!-\!y)f_2^{\prime}-(D\!+\!1)f_2\big]
   \Big\}
\,.
\nonumber
\end{eqnarray}
Note that subtracting
(\ref{tensor structure:3}) from~(\ref{tensor structure:2}) yields
\begin{eqnarray}
R^{\mu\nu}
     \{(\partial_{\nu}\partial^\prime_{\alpha} y)\times f_1(y)
       +(\partial_{\nu}y)\times(\partial^\prime_{\alpha} y)\times f_2(y)\}
 &=& \, H^2(\partial^{\mu}\partial^{\prime}_{\alpha}y)
     \big[(D\!-\!1)f_1\big]
      + H^2(\partial^{\mu}y)\times(\partial^{\prime}_{\alpha}y)
          \big[(D\!-\!1)f_2\big]
\,,\quad
\label{Ricci tensor:action}
\end{eqnarray}
as it should (see Eq.~(\ref{photon operator Lmn})).
Here we made use of
$R^{\mu\nu}V_\nu=(\nabla^\nu\nabla^\mu-\nabla^\mu\nabla^\nu)V_\nu$ for any
vector field $V_\nu$.

 Now upon inserting
the results~(\ref{Box on propagator}--\ref{tensor structure:3})
into the photon propagator equation~(\ref{photon propagator:eom}) we get,
\begin{eqnarray}
&&  L^{\mu\nu}\{(\partial_{\nu}\partial^\prime_{\alpha} y)\times f_1(y)
   + (\partial_{\nu}y)\times(\partial^\prime_{\alpha} y)\times f_2(y)\}
\label{L on propagator:tensor structure} \\
   &&\hskip 3cm
= \, H^2(\partial^{\mu}\partial^{\prime}_{\alpha}y)
\Big\{\Big[
      (4y\!-\!y^2)f_1^{\prime\prime}
       +\Big((D\!-\!1)+\frac{1}{\xi}\Big)(2\!-\!y)f_1^{\prime}-\frac{D}{\xi}f_1
     \Big]
\nonumber\\
   &&\hskip 5.2cm
+ \, \Big[\!-\!\Big(1-\frac{1}{\xi}\Big)(4y\!-\!y^2)f_2^{\prime}
         -\Big(D\!-\!1-\frac{D\!+\!1}{\xi}\Big)(2\!-\!y)f_2
      \Big]\Big\}
\qquad
 \nonumber \\
&&\hskip 3cm
 +\,H^2(\partial^{\mu}y)(\partial^{\prime}_{\alpha}y)
 \Big\{\Big[\!-\!\Big(1-\frac{1}{\xi}\Big)(2\!-\!y)f_1^{\prime\prime}
+\Big(D\!-\!1-\frac{D\!+\!1}{\xi}\Big)f_1^{\prime}\Big]
\nonumber\\
&&\hskip 5.5cm
  +\,\Big[\frac{4y\!-\!y^2}{\xi}f_2^{\prime\prime}
  +\Big(1+\frac{D\!+\!3}{\xi}\Big)(2\!-\!y)f_2^{\prime}
   -\Big(D\!-\!1+\frac{D\!+\!1}{\xi}\Big)f_2
  \Big]\Big\}
\nonumber\\
&&\hskip 3.cm
  =\, (\partial^\mu\partial^{\prime}_{\alpha}y)
         \frac{\imath\delta^D(x\!-\!x')}{\sqrt{-g}(-2H^2)}
\,.
\end{eqnarray}
This result is used in the main text to obtain
Eqs.~(\ref{scalar equation:1}--\ref{scalar equation:2})
for the scalar structure functions $f_1$ and $f_2$.


\section{An alternative method to calculate the scalar structure function $A_2$}
\label{An alternative method to calculate the scalar structure function A_2}

This appendix is an attempt to construct in an alternative manner
the second scalar structure function $A_2$ 
defined in Eq.~(\ref{propagator: tensor structures a}).
Namely, we shall attempt to solve for $A_2$ by performing the suitable double
integral in Eq.~(\ref{eom:A2:I}). For simplicity, we work here in $D=4$.
Taking account of the four dimensional form for $A_1$ and $I[A_1]$ 
in Eqs.~(\ref{intA1D4}--\ref{A1D4}) 
and the corresponding source function $s_\xi$ in Eq.~(\ref{sxiD4}),
it is straightforward to perform the integrals in~(\ref{eom:A2:I}). 
The (na\"\i ve) result is (up to a constant), 
\begin{eqnarray}
 \tilde A_2(y) &=& \frac{1}{8\pi^2}\bigg\{(1\!-\!\xi)
          \bigg[\frac13\frac{1}{4\!-\!y}+\frac12\ln\Big(\frac{y}{4}\Big)
               -\frac16\ln\Big(1\!-\!\frac{y}{4}\Big)
          \bigg]
\nonumber\\
&&\hskip 0.9cm
 -\,\big(3\!-\!\xi\big)
   \bigg[\!-\!\frac59\frac{1}{4\!-\!y}+\frac{y}{6(4\!-\!y)}\ln\Big(\frac{y}{4}\Big)
     +\frac{5}{18}\ln\Big(1\!-\!\frac{y}{4}\Big)
     +\frac{1}{3}{\rm Li}_2\Big(1\!-\!\frac{y}{4}\Big)
   \bigg]
\bigg\}
\, .
\label{Appx C:A2}
\end{eqnarray}
This result does not reproduce the right singular structure of Eq.~(\ref{eom:A2:I}).
This can be seen from the near pole (analytic) structure of~(\ref{Appx C:A2}), which 
around $y\sim 4$ and $y\sim 0$ is 
\begin{eqnarray}
\tilde A_2(y\sim 4) &=& \frac{1}{8\pi^2}
\bigg[\frac{1\!-\!\xi}{3}+\frac{5(3\!-\!\xi)}{9}\bigg]
          \bigg[\frac{1}{4\!-\!y}-\frac{1}{2}\ln\Big(1\!-\!\frac{y}{4}\Big)\bigg]
+{\cal O}\Big((4\!-\!y)^0\Big)
\nonumber\\
\tilde A_2(y\sim 0) 
   &=& \frac{1}{8\pi^2}\bigg\{\frac{1\!-\!\xi}{2}\ln\Big(\frac{y}{4}\Big)
\bigg\} 
  + {\cal O}\Big((4\!-\!y)^0\Big)
\,,
\label{Appx C:sxi:sg}
\end{eqnarray}
such that, when the operator 
$\Box/H^2 = (4y\!-\!y^2)(d/dy)^2+4(2\!-\!y)(d/dy)$ 
acts on $A_2$ and one takes proper account of the singular (pole) structure
indicated by the $\imath\epsilon$ prescription in $y=y^{++}(x;x^\prime)$,
one gets from~(\ref{Appx C:A2}--\ref{Appx C:sxi:sg}),
\begin{equation}
\frac{\Box}{H^2}\tilde A_2(y(x;x^\prime)) = s_\xi(y(x;x^\prime)) 
 -\frac12 \bigg[\frac{1\!-\!\xi}{3}+\frac{5(3\!-\!\xi)}{9}\bigg]
                  \frac{\imath \delta^4(x\!-\!\bar x^\prime)}{H^4\sqrt{-g}}
\,,
\label{Appx C:A2:sg structure}
\end{equation}
where we took account of (see Eq.~(\ref{secondDis4})),
\begin{equation}
\Box \imath G_{A_2}(x;x') 
    =\frac{\imath \delta^4(x\!-\! x^\prime)+\imath \delta^4(x\!-\!\bar x^\prime)}
          {\sqrt{-g}}
\,,
\label{Box na GA2}
\end{equation}
where $\bar x^\mu =(-\eta,\vec x\,)$ is the antipodal point of $x^\mu$.
Adding a homogeneous, de Sitter invariant,
solution $\propto G_{A_2}$ to $A_2$ cannot remove 
the singular contribution in Eq.~(\ref{Appx C:A2:sg structure}). 
Indeed, from~(\ref{Box na GA2}) it immediately follows that adding such a term
can only replace the delta function
at the antipodal point with that at the light cone $x^\mu=x^{\prime\mu}$,
such that we have
\begin{eqnarray}
A_2 &=& \tilde A_2 
      + \frac{1}{2H^2}\bigg[\frac{1\!-\!\xi}{3}+\frac{5(3\!-\!\xi)}{9}\bigg]
             \imath G_{A_2}(x;x^\prime)
\nonumber\\
\frac{\Box}{H^2}A_2(y(x;x^\prime)) &=& s_\xi(y(x;x^\prime)) 
 +\frac12 \bigg[\frac{1\!-\!\xi}{3}+\frac{5(3\!-\!\xi)}{9}\bigg]
                  \frac{\imath \delta^4(x\!-\!x^\prime)}{H^4\sqrt{-g}}
\,.
\label{Appx C:A2:sg structure:2}
\end{eqnarray}
Even though this new choice for $A_2$ has the right (Hadamard) singular structure, 
the resulting photon propagator~(\ref{propagator: tensor structures a}) does not
satisfy the correct equation of motion~(\ref{photon propagator:eom}), 
that is for this propagator the delta function structure on the right hand side 
of~(\ref{photon propagator:eom}) is incorrect.
This means that it is impossible to construct a de Sitter invariant 
solution to $A_2$ of the form $A_2(y(x;x'))$ with the right singular structure,
which is at odds with the statement made in Ref.~\cite{Allen:1986}.
Curiously, the singular terms in Eqs.~(\ref{Appx C:A2:sg structure})
and~(\ref{Appx C:A2:sg structure:2}) vanish
in the gauge $\xi = 9/4$, which differs from the Feynman gauge, $\xi = 1$, 
used in Ref.~\cite{Allen:1986}. 
For simplicity we have here considered the $D=4$ case, 
but of course the same conclusion can be reached for a general number 
of space-time dimensions.


\end{document}